\documentclass[12pt,a4paper,final]{iopart}
\usepackage{iopams}  
\usepackage{graphicx}
\usepackage[breaklinks=true,colorlinks=true,linkcolor=blue,urlcolor=blue,citecolor=blue]{hyperref}
\usepackage{siunitx}
\newcommand{\hightc}{high-$T_c$}

\newcommand{\lowtc}{low-$T_c$}

\newcommand{\rhz}{$\sqrt{\textrm{Hz}}$}
\newcommand{\uv}{\si{\micro\volt}}
\newcommand{\um}{\si{\micro\meter}}
\newcommand{\uA}{\si{\micro\ampere}}
\newcommand{\uphi}{\si{\micro}$\Phi_0$}
\newcommand{\uvphi}{\uv/$\Phi_0$}
\newcommand{\uphihz}{\uphi/\rhz}
\newcommand{\Vphi}{$V_{\Phi}$}
\newcommand{\Sphi}{$S_\Phi^{1/2}$}
\newcommand{\Sv}{$S_V^{1/2}$}
\newcommand{\Sb}{$S_B^{1/2}$}
\newcommand{\Aeff}{$A_{eff}$}

\begin{document}

\title[The role of kinetic inductance on the performance of YBCO SQUID magnetometers]{The role of kinetic inductance on the performance of YBCO SQUID magnetometers}

\author{S Ruffieux$^{1}$, A Kalaboukhov$^{1}$, M Xie$^{1}$, M Chukharkin$^{2}$, C Pfeiffer$^{1}$, S Sepehri$^{1}$, J F Schneiderman$^{3}$ and D Winkler$^{1}$}
\address{$^1$Department of Microtechnology and Nanoscience -- MC2, Chalmers University of Technology, SE-41296 Gothenburg, Sweden}
\address{$^2$Chalmers Industriteknik, SE-41288 Gothenburg, Sweden}
\address{$^3$MedTech West and the Institute of Neuroscience and Physiology, University of Gothenburg, SE-40530 Gothenburg, Sweden}
\ead{ruffieux@chalmers.se}

\begin{abstract}
Inductance is a key parameter when optimizing the performance of superconducting quantum interference device (SQUID) magnetometers made from the high temperature superconductor YBa$_2$Cu$_3$O$_{7-x}$ (YBCO) because lower SQUID inductance $L$ leads to lower flux noise, but also weaker coupling to the pickup loop.
In order to optimize the SQUID design, we combine inductance simulations and measurements to extract the different inductance contributions, and measure the dependence of the transfer function \Vphi{} and flux noise \Sphi{} on $L$. A comparison between two samples shows that the kinetic inductance contribution varies strongly with film quality, hence making inductance measurements a crucial part of the SQUID characterisation.
Thanks to the improved estimation of the kinetic inductance contribution, previously found discrepancies between theoretical estimates and measured values of \Vphi{} and \Sphi{} could to a large extent be avoided.
We then use the measurements and improved theoretical estimations to optimize the SQUID geometry and reach a noise level of \Sb{} = 44 fT/\rhz{} for the best SQUID magnetometer with a 8.6 mm $\times$ 9.2 mm directly coupled pickup loop.
Lastly, we demonstrate a method for reliable one-time sensor calibration that is constant in a temperature range of several kelvin despite the presence of temperature dependent coupling contributions, such as the kinetic inductance.
The found variability of the kinetic inductance contribution has implications not only for the design of YBCO SQUID magnetometers, but for all narrow linewidth SQUID-based devices operated close to their critical temperature.
\end{abstract}


\section{Introduction}
Superconducting quantum interference device (SQUID) magnetometers and gradiometers made from the high critical temperature (\hightc{}) superconducting material YBa$_2$Cu$_3$O$_{7-x}$ (YBCO) are nowadays used in various applications like geophysical exploration \cite{Leslie2008,Hato2013,Chwala2015}, nondestructive evaluation (NDE) and contaminant detection \cite{Hatsukade2005,Faley2017b,Tanaka2017}, as well as in biomedical applications, such as magnetocardiography (MCG) \cite{Yang2009}, magnetoencephalography (MEG) \cite{Oeisjoeen2012,Faley2017a,Schneiderman2019}, and biosensing using magnetic nanoparticles \cite{Enpuku2017,Sepehri2018,Tanaka2018}.
Thanks to their high critical temperature, \hightc{} SQUIDs have reduced cooling requirements compared to their \lowtc{} counterparts, which allows for cheaper sensor operation, more compact systems \cite{Chwala2015}, and reduced sensor standoff distance to nearby sources leading to higher signal amplitudes \cite{Oeisjoeen2012,Xie2016,Schneiderman2019}.

However, operation of the SQUIDs at temperatures around the boiling point of liquid nitrogen comes with a significant amount of thermal noise that degrades the flux-to-voltage transfer function \Vphi{} rapidly with increasing SQUID inductance $L$ \cite{Enpuku1993, Enpuku1994}. Low flux noise \Sphi{} = \Sv{}/\Vphi{} can be achieved by decreasing $L$. However, in order to make a sensitive magnetometer or gradiometer, a pickup loop needs to be coupled to the SQUID inductance. As the coupling increases with inductance, there is a trade-off between low flux noise and strong coupling when optimizing the sensor noise performance.

Previous reports of discrepancies between theoretical estimates and measured values both for \Vphi{} and \Sphi{} \cite{Enpuku1995,Koelle1999,Blomgren2002,Lam2013} complicate the magnetometer optimization process. The theoretical estimates depend on the SQUID critical current $I_c$, the normal resistance $R_n$, and the SQUID inductance $L$ \cite{Tesche1977,Enpuku1993,Koelle1999}. While the junction parameters $I_c$ and $R_n$ are generally determined for every SQUID from its current-voltage characteristic ($I$-$V$ curve), the SQUID inductance is typically calculated numerically and assumed to be constant for the device design used. Calculation of the kinetic inductance contribution to $L$ requires knowledge of the London penetration depth $\lambda$, which strongly varies with the critical temperature $T_c$ of the YBCO film and the operation temperature $T$ when operating the device close to $T_c$ \cite{Lee1993,Brake1997a}:
\begin{equation}
\lambda = \frac{\lambda_0}{\sqrt{1-\left(T/T_c\right)^2}},
\label{eq:lambda}
\end{equation}
where $\lambda_0$ is the London penetration depth at 0 K.
An error in the estimation of $L$ due to kinetic inductance is hence a possible reason for the discrepancies found \cite{Enpuku1995}. Other possible reasons are improperly set bias conditions, environmental or electronics noise, excess currents, resonances, or asymmetries in the junction parameters  \cite{Enpuku1995,Koelle1999,Tesche1977}.

Kinetic inductance measurements in YBCO dc SQUIDs have been performed for various SQUID designs, e.g., washer type SQUID magnetometers \cite{Grundler1995}, hairpin SQUIDs with a ground plane \cite{Terai1997}, biepitaxial SQUIDs (to measure the crystal orientation dependence of $\lambda$ in YBCO) \cite{Johansson2009}, nanoSQUIDs (that are known to have high kinetic inductance contributions due to their small dimensions) \cite{Schwarz2013,Schwarz2015}, and most recently nano-slit SQUIDs \cite{Li2019}. While kinetic inductance contributions are thus reportedly significant, the effects of film quality and sample-to-sample variation remain an open question.

In this paper, we combine inductance simulations and measurements to study the different inductance contributions in single layer YBCO hairpin dc SQUID magnetometers with a directly coupled pickup loop, which are used in our 7-channel on-scalp MEG system \cite{Pfeiffer2019}. We present measurements from 2 samples (10 bare SQUIDs) with slightly different film quality to show that the kinetic inductance contributions can differ strongly. The results show the importance of inductance measurements to significantly reduce the error in the estimation of $L$.

We then investigate the role of kinetic inductance on the sensor performance characterized by \Vphi{}, \Sphi{}, the coupling described by the coupling inductance $L_c$ or the effective area $A_{eff}$, and ultimately the magnetic field noise \Sb{}. The dependence of \Vphi{} and \Sphi{} on the measured inductance has been examined before \cite{Mitchell2002,Mitchell2003}, however, it is difficult to use these measurements to optimize L as the flux noise was dominated by large low frequency noise.
To avoid this problem, we operate our SQUIDs in a flux-locked loop (FLL) with AC bias reversal to cancel critical current fluctuations \cite{Drung2003}.

Finally, measurements of the magnetometer effective area $A_{eff}$ as a function of temperature showed that the coupling is temperature dependent due to the kinetic inductance contribution to $L_c$ \cite{Brake1997a,Grundler1995}. Sensor operation temperature fluctuations could thus pose an experimental challenge in terms of flux-to-field calibration. We therefore include measurements of the temperature dependence of the magnetometer coupling in the temperature range of interest and present a method to achieve temperature independent magnetometer calibration.

The inductance optimization is here performed for magnetometers with a directly coupled pickup loop, but it is straightforward to extend the results to gradiometers as well. The measured inductance variation between samples furthermore has implications for the design of all kinds of YBCO SQUID-based devices operating close to $T_c$.

\section{Methods}
\subsection{Magnetometer design and optimization}
\begin{figure}[hbt!]
 \centering
 \includegraphics[width=1\linewidth]{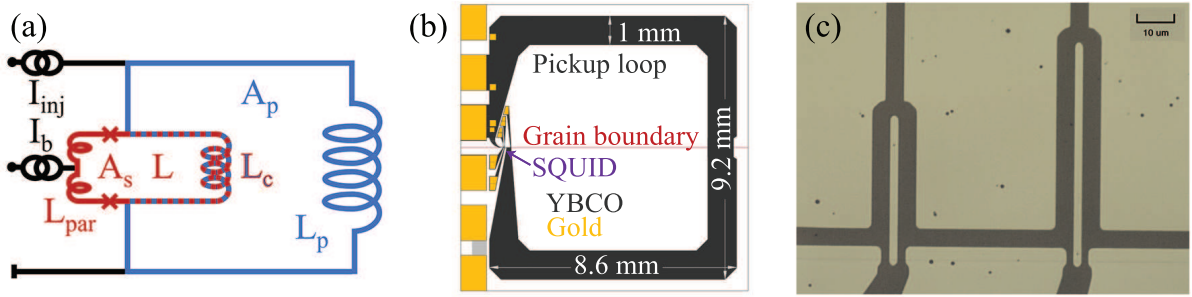}
 \caption{Hairpin dc SQUID magnetometer with a directly coupled pickup loop. (a) Equivalent circuit showing the dc SQUID (small loop, red) and the pickup loop (big loop, blue). The dc SQUID has two Josephson junctions shown as crosses, an effective area $A_s$ and an inductance $L$ consisting of the coupling inductance $L_c$ and the parasitic inductance $L_{par}$. The pickup loop has an effective area $A_p$ and an inductance $L_p$ that includes the coupling inductance $L_c$ to the SQUID loop. (b) CAD design of a magnetometer with a 1 mm linewidth pickup loop made on a 10 mm $\times$ 10 mm substrate. (c) Micrograph of the SQUID area showing two (redundant) YBCO SQUIDs with narrow linewidth.}
 \label{fig:Magnetometer}
\end{figure}

The magnetometer design consists of a hairpin dc SQUID directly coupled to a pickup loop as shown in Figure \ref{fig:Magnetometer}. This sensor design is beneficial as the complete magnetometer can be made from a single layer YBCO film, thus avoiding the challenge of fabricating low noise multilayer YBCO structures or having to assemble flip-chip devices \cite{Lee1995,Faley2017a,Kaczmarek2018}. Furthermore, the coupling of low frequency flux noise into the SQUID due to moving vortices in the pickup loop can be minimized by using narrow linewidths ($\sim$4 \um) for the SQUID loop and current injection lines, as well as locating the SQUID at a sufficient distance ($\sim$100 \um) from the solid pickup loop \cite{Cho1999,Du2004}.

Optimization of the magnetometer's magnetic field noise \Sb{} involves accounting for the flux noise and the effective area \Aeff{} of the magnetometer: \Sb{} = \Sphi{}/\Aeff{}. The effective area of a dc SQUID magnetometer with a directly coupled pickup loop with inductance $L_p$ and effective area $A_p$ can be approximated by 
\begin{equation}
    A_{eff} = A_s + L_c \cdot \frac{A_p}{L_p}  \approx L_c \cdot \frac{A_p}{L_p}
    \label{eq:Aeff}
\end{equation}
for negligible SQUID effective area $A_s$. The coupling inductance $L_c$ between the pickup loop and the SQUID loop is determined by the segment shared by the two loops. The magnetic field noise can thus be divided into a SQUID dependent factor (\Sphi{}/$L_c$) and a pickup loop dependent factor ($L_p/A_p$), meaning that the SQUID and the pickup loop can be optimized individually. We hence begin our optimization by studying bare hairpin SQUIDs first. Then we select the best SQUID design and make a complete magnetometer.

\subsection{Sample fabrication}
\begin{figure}[tbh!]
 \centering
 \includegraphics[width=1\linewidth]{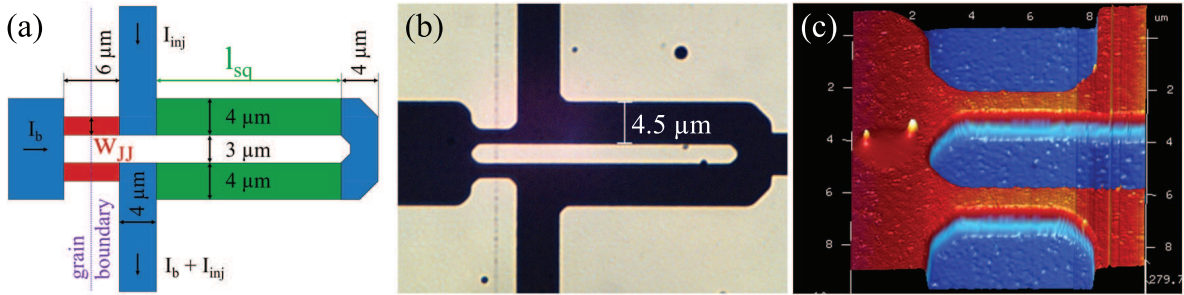}
 \caption{Hairpin SQUID design (a) CAD design of the hairpin SQUID showing the dimensions and  applied currents. We vary the length $l_{sq}$ of the SQUID loop from 10 \um{} to 50 \um{}, and the width $w_{JJ}$ of the Josephson junctions that are formed where the YBCO film crosses the grain boundary around 1 \um{}. (b) Backlight micrograph of a fabricated hairpin SQUID showing the grain boundary. The resulting YBCO linewidth is 0.5 \um{} wider than in the design. (c) AFM image of the Josephson junction area of a SQUID with $w_{JJ}$ = 1 \um{}.}
 \label{fig:Design}
\end{figure}

We fabricated two chips with bare hairpin SQUIDs to study the different inductance contributions and how the inductance influences \Vphi{} and \Sphi{}.
The main difference between the two samples regarding fabrication is that the YBCO film was directly grown on the STO substrate for sample A, while a CeO$_2$ buffer layer was used in sample B.
The SQUID design with the relevant dimensions is shown in Figure \ref{fig:Design}a. For both samples, the length of the SQUID loop $l_{sq}$ was varied from 10 \um{} to 50 \um{} in steps of 10 \um{} in order to change the coupling inductance \cite{Mitchell2002,Mitchell2003}. For the junction width $w_{JJ}$, we aimed at different sizes around 1 \um{} as previous SQUIDs made from YBCO films grown directly on STO substrates showed that such narrow junctions were necessary to achieve SQUID critical currents below 80 \uA{}.

The two samples were made on STO bicrystal substrates from the same batch with a misorientation angle of 22.6\si{\degree} (Shinkosha, Japan). The 140 nm thick YBCO films were deposited with pulsed laser deposition (PLD) using an excimer laser of 248-nm wavelength. For sample A, the YBCO film was grown directly on the STO bicrystal substrate using the optimized deposition parameters given in Table \ref{tab:PLD}. For sample B, a 50 nm thick CeO$_2$ buffer layer was grown first using RF sputtering. The YBCO deposition parameters were reoptimized for the growth of YBCO on CeO$_2$ and can also be found in Table \ref{tab:PLD}. The YBCO film on sample B was grown following the CeO$_2$ deposition without breaking the vacuum. The fabrication process after PLD was the same for both samples: in the next step the YBCO films were protected by a 50 nm thick in-situ sputtered gold layer.

\Table{\label{tab:PLD} YBCO PLD deposition parameters}
\br
 & Sample A & Sample B \\
\mr
Deposition temperature & 750 \si{\degree}C & 750 \si{\degree}C \\
Deposition pressure & 1.6 mBar & 0.6 mBar \\
Distance to target & 52.5 mm & 54 mm \\
Laser energy density & 1.58 J/cm$^2$ & 1.5 J/cm$^2$ \\
Pulse frequency & 5 Hz & 5 Hz \\
Number of pulses & 2000 & 2000 \\
Post annealing pressure & 850 mBar & 0.6 mBar\\
\br
\endTable

The SQUIDs were patterned using a hard carbon hard mask and argon ion milling to achieve $\lesssim$ 1 \um{} junctions with well defined edges as shown in Figure \ref{fig:Design}c. The patterning procedure is based on a chromium layer and oxygen plasma etching to define the carbon mask \cite{Arpaia2013}. A laser writer was used to expose the pattern.
The argon ion etching with an ion beam voltage of 300 V and a current density of 0.08 mA/cm$^2$ was monitored by secondary ion mass spectrometry (SIMS) for endpoint detection. After ion milling, the carbon mask was removed with oxygen plasma stripping, and gold contact pads were defined in a lift-off process. Finally, the capping gold layer was removed in a short (4 min) argon ion etch.

\begin{figure}[htb!]
 \centering
 \includegraphics[width=0.6\linewidth]{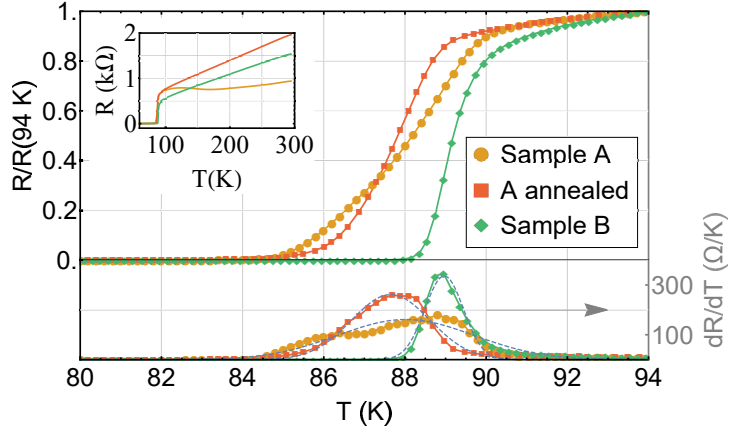}
 \caption{\label{fig:Tc} Normalized resistance versus temperature measurements showing the superconducting transition. The inset shows the full temperature range and that a linear temperature dependence above the transition is also obtained for sample A after annealing. The temperature derivative $dR/dT$ is plotted in the lower part of the figure. We define the critical temperature $T_c$ as the maximum of the Gaussian fit (dashed line) to $dR/dT$, and its FWHM as the width of the transition.}
\end{figure}

Resistance versus temperature measurements of both fabricated samples are shown in Figure \ref{fig:Tc}. As sample A was underdoped, we annealed it at 600 \si{\degree}C in 650 Torr oxygen pressure for 2 hours. The resulting curve showed the normal linear temperature dependence with a sharper transition (width 2.3 K instead of 3.9 K) at a slightly lower critical temperature (87.7 K instead of 88.0 K). Sample B has a very sharp transition with a width of 1.1 K and a higher critical temperature of 89.0 K.

Based on the results from the inductance measurements and bare SQUID optimization performed for samples A and B (more on this below), we furthermore fabricated new samples containing magnetometers with directly coupled pickup loops. For each magnetometer, two hairpin SQUIDs with $l_{sq}$ = 30 \um{} and 50 \um{} (see Figure \ref{fig:Magnetometer}c) were coupled to the pickup loop shown in Figure \ref{fig:Magnetometer}b. These magnetometers were fabricated in the same way as sample B and are used here for noise and effective area measurements.

\subsection{Inductance simulation}
The different inductance contributions were extracted by numerically solving the London and Maxwell equations in the COMSOL Multiphysics software (COMSOL Inc., Stockholm, Sweden) using the stream function formalism established by Khapaev \cite{Khapaev1997}.
In this approach, the system is treated as 2-dimensional under the conditions that the film thickness t fulfills $t \ll \lambda$ and $t \ll l$, where $l$ is the characteristic length of the structure -- both of which are fulfilled in our case.
The thickness dependence is then described by the Pearl penetration length $\lambda_p = \lambda^2/t$.
By calculating the total energy of the system for different current boundary conditions, it is possible to extract $L$, $L_c$, and $L_p$ of our magnetometer \cite{Xie2017}.
This simulation tool has been used successfully for devices with high kinetic inductance contributions including nanowire-based SQUIDs \cite{Arzeo2016,Xie2017} and biepitaxial SQUIDs \cite{Johansson2009}.

We can differentiate between kinetic and geometric inductance by calculating the current energy and the magnetic field energy separately. However, in order to calculate the kinetic inductance, knowledge of $\lambda_p$ is necessary. We pick $\lambda_p$ = 800 nm for the simulation based on the common values $\lambda$ = 400 nm and $t$ = 200 nm. As the
kinetic inductance is proportional to $\lambda_p$, while the geometric inductance is independent of $\lambda_p$ \cite{Khapaev1997}, the real inductances can be calculated from the simulated values once $\lambda_p$ is determined.

For the simulation, we use the SQUID design shown in Figure \ref{fig:Design}a. However, since the linewidth of the fabricated devices is $\sim$0.5 \um{} wider than in the design (see Figure \ref{fig:Design}b), we adjusted the geometry of the model to match that of the actual devices.
The SQUID is coupled to the 1 mm wide pickup loop shown in Figure \ref{fig:Magnetometer}b, which has an inductance $L_p$ = 17.88 nH + $L_c$. The detailed design of the pickup loop does not affect the results for $L_c$ and $L$, but having a pickup loop is necessary to extract $L_c$.

\subsection{Measurement methods}
\begin{figure}[htb!]
 \centering
 \includegraphics[width=0.6\linewidth]{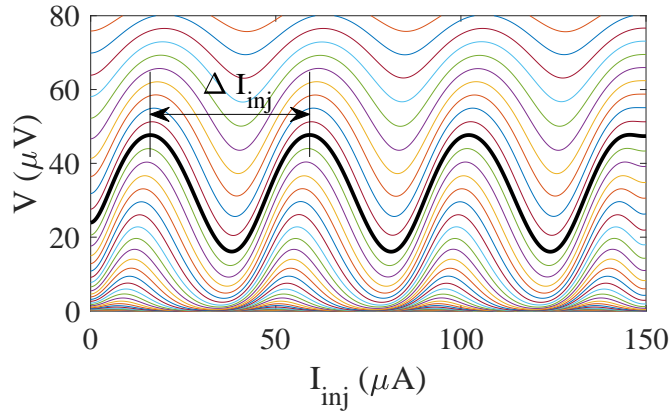}
 \caption{\label{fig:IVPhi} SQUID voltage $V$ as a function of injected current $I_{inj}$ (effectively flux bias) for different bias currents. The curve with the highest voltage modulation amplitude $\Delta V$ is marked in black. The voltage modulation period $\Delta I_{inj}$ is used to extract the coupling inductance $L_c$.}
\end{figure}

All measurements were performed inside a magnetically shielded room. We used a direct readout dc SQUID electronics SEL-1 (Magnicon GmbH, Hamburg, Germany). For the characterization and coupling inductance measurements, the SQUIDs were cooled inside a dipstick (filled with 0.8 bar helium exchange gas) immersed in liquid nitrogen; the resulting operation temperature was $\sim$78 K.
To couple flux into the SQUID and to measure $L_c$, we directly injected the current $I_{inj}$ into the SQUID loop as illustrated in Figure \ref{fig:Design}a. An example of the resulting voltage modulation for different bias currents is shown in Figure \ref{fig:IVPhi}. We extracted $L_c$ from the voltage modulation period $\Delta I_{inj}$ using the relation $\Delta I_{inj}\cdot L_c = \Phi_0$ \cite{Tesche1977}.

The maximal transfer function \Vphi{} was obtained from the slope of the $V$-$I_{inj}$ curves. These curves were also used to get the $I$-$V$ curve with zero applied flux in the SQUID, from which the SQUID critical current $I_c$ and normal resistance $R_n$ were obtained. Several curves included characteristics of excess currents $I_{ex}$, which are defined by the current axis intercept of a linear fit to the $I$-$V$ curve at large currents \cite{Tafuri2005}. As excess currents are not described with the resistively shunted junction (RSJ) model used to simulate the SQUID behaviour \cite{Tesche1977,Enpuku1993,Koelle1999}, we replace $I_c$ with the reduced SQUID critical current $I_c^* = I_c - I_{ex}$ when comparing our results with theoretical predictions.

Flux noise measurements were performed with the sample inside a superconducting shield using an FFT spectrum analyzer (Keysight Dynamic Signal Analyzer). The SQUIDs were operated in a flux-locked loop with AC bias reversal at 40 kHz to cancel critical current fluctuations. White noise levels were determined by averaging 50 noise spectra and then averaging the noise between 1 kHz and 10 kHz.

The effective area of the magnetometer was obtained by applying a known magnetic field with a calibrated Helmholtz coil. We varied the amplitude of the applied field and linearly fitted the output flux measured by the SQUID in FLL-mode.

In order to be able to vary the sensor operation temperature, we placed the magnetometer inside a liquid nitrogen cryostat instead of the dipstick. The cryostat temperature can be controlled by pumping on the liquid nitrogen bath, which reduces the boiling point. The temperature as a function of pumping pressure was calibrated with a diode temperature sensor in  a separate cool down (the temperature sensor introduces measurement noise).

\section{Results}

\subsection{Inductance simulation}
\begin{figure}[tbh!]
 \centering
 \includegraphics[width=0.7\linewidth]{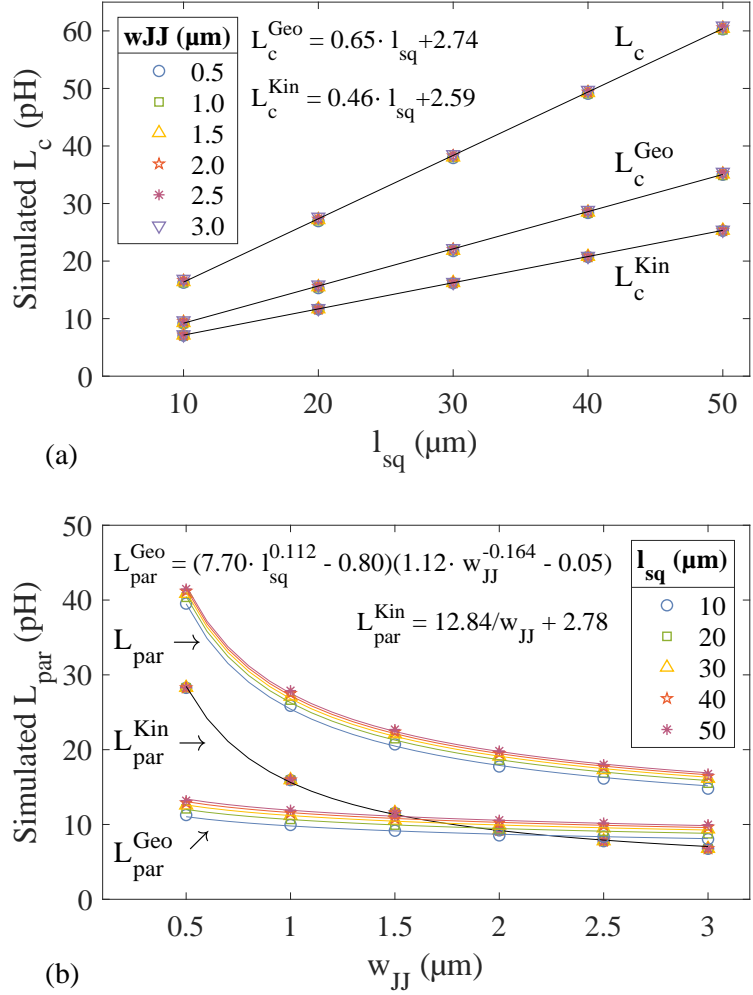}
 \caption{\label{fig:Simulation} Simulated SQUID coupling inductance $L_c$ (a) and parasitic inductance $L_{par}$ (b) as a function of SQUID loop length $l_{sq}$ and Josephson junction width $w_{JJ}$. In both cases the total inductance is made up of a geometric and a kinetic inductance contribution. The best fits are shown by lines and their functions are given in the figure. For the kinetic inductance contribution we use $\lambda_p = 800$ nm.}
\end{figure}

The results of the inductance simulation are plotted in Figure \ref{fig:Simulation}. Best fits to the simulated inductances are indicated with solid lines and are later used to calculate the inductance contributions in our measured SQUIDs.

For the coupling inductance, both the geometric ($L_c^{Geo}$) and the kinetic ($L_c^{Kin}$) terms are independent of the Josephson junction width $w_{JJ}$ and scale linearly with $l_{sq}$. Varying $l_{sq}$ is hence an effective way to adjust $L_c$. Furthermore, it is clear that the kinetic inductance is not negligible as it accounts for around 42\% of $L_c$ for $\lambda_p = 800$ nm.
Both the slopes of $L_c^{Geo}$ and $L_c^{Kin}$ as a function of $l_{sq}$ match predictions from analytical formulas well. The geometrical inductance per unit length of a thin coplanar stripline with spacing s between the strips of width w is $\mu_0 K(k)/K(k')$, where $\mu_0$ is the vacuum permeability and $K(k)$ is the complete elliptic integral of the first kind
with a modulus $k=s/(s+2w)$ and $k' = \sqrt{1-k^2}$ \cite{Yoshida1992}. For our simulated SQUID with $s = 2.5$ \um{} and $w = 4.5$ \um{}, the analytical formula predicts a slope of 0.64 pH/\um{} in good agreement with the simulated slope of 0.65 pH/\um{}. The kinetic inductance of a strip with linewidth $w$, thickness $t$ and length $l$ is \cite{Lee1993}:
\begin{equation}
L_{strip}^{kin} = \mu_0 \frac{l}{wt}\lambda^2 = \mu_0 \frac{l}{w}\lambda_p.
\label{eq:Lkin}
\end{equation}
This formula predicts a slope of 0.45 pH/\um{} for the kinetic inductance contribution to $L_c$ (using $w = 4.5$ \um{} and $\lambda_p = 800$ nm), which is also in good agreement with the simulated value of 0.46 pH/\um{}.

The parasitic inductance depends strongly on $w_{JJ}$ and increases substantially for submicron junctions due to their large parasitic kinetic inductance contribution $L_{par}^{Kin}$.
To achieve low $L_{par}^{Kin}$, junctions with a large cross section $w_{JJ}\cdot t$ and short bridges (i.e., the strips colored red in Figure \ref{fig:Design} where the Josephson junctions are created) are favourable as predicted by equation (\ref{eq:Lkin}). However, the junction cross section is dictated by the targeted critical current, and the minimal bridge length by the alignment precision possible during fabrication. Hence the minimal achievable $L_{par}$ is strongly related to the chosen junction technology, as well as $\lambda_p$.

\subsection{SQUID characterization and inductance measurements}
For the following measurements and analysis we selected 5 SQUIDs with different $l_{sq}$ from each of the two samples. The parameters of these 10 SQUIDs are summarized in Table \ref{tab:SQUIDs}. When selecting the SQUIDs, we aimed for high $w_{JJ}$ (giving the lowest $L_{par}$) under the condition that $I_c$ \textless 80 \uA{}. The latter is a practical limitation of the SQUID electronics we use, whose bias current range is +/- 250 \uA{}, and follows from the recommendation that the bias current is measured up to at least 3$I_c$ to allow good fitting of the $I$-$V$ curve.
A notable difference between the samples is that the SQUIDs on sample A have a higher junction critical current density (average $J_c = 2.1\cdot 10^4$ A/cm$^2$) than those on sample B (average $J_c = 6.5\cdot 10^3$ A/cm$^2$), as well as higher excess currents $I_{ex}$ (5-24\% and $<$ 2\% of $I_c$ for samples A and B, respectively). We see the same behaviour when comparing other samples with and without a CeO$_2$ buffer layer, and hence attribute this change in junction properties to the buffer layer. The $I_c^*R_n$ products are similar in both samples with an average value of 108 \uv{}.

\fulltable{\label{tab:SQUIDs} SQUID parameters at $T \approx$ 78 K. Device names refer to the sample (A or B) and $l_{sq}$ in \um{}.}
\br
Name & $l_{sq}$ & $w_{JJ}$ & $I_c$ ($I_c^*$) & $R_n$ & $I_cR_n$ ($I_c^*R_n$) & $L_c$ & $\lambda_p$ & $L$ & $\Delta V$ & \Vphi{} &\Sphi{} &\Sphi{}/$L_c$\\
  & (\um{}) & (\um{}) & (\uA{}) & ($\Omega$) & (\uv{}) & (pH) & (\um) & (pH) & (\uv{}) & ($\frac{\uv}{\Phi_0}$) & ($\frac{\si{\micro}\Phi_0}{\sqrt{\textrm{Hz}}}$) & ($\frac{\si{\micro}\Phi_0}{\sqrt{\textrm{Hz}}\cdot \textrm{nH}}$)\\
\mr
A10 & 10 & 0.9 & 42 (32) & 3.1 & 129 (98) & 33.1 & 2.7 & 100 & 18 & 65 & 11.0 & 332\\
A20 & 20 & 1.1 & 56 (51) & 1.7 & 95 (86) & 46.5 & 2.1 & 95 & 11 & 37 & 18.5 & 398\\
A30 & 30 & 1.2 & 70 (67) & 2.0 & 139 (133) & 75.2 & 2.6 & 130 & 13 & 50 & 15.6 & 207\\
A40 & 40 & 1.0 & 73 (56) & 1.7 & 122 (93) & 83.1 & 2.1 & 136 & 9 & 32 & 17.6  & 211\\
A50 & 50 & 0.8 & 48 (46) & 2.4 & 115 (110) & 128.5 & 3.0 & 210 & 5 & 17 & 31.5 & 245\\
\mr
B10 & 10 & 1.3 & 19 (18) & 5.5 & 103 (100) & 15.3 & 0.7 & 35 & 55 & 195 & 2.6 & 167\\
B20 & 20 & 1.4 & 26 (25) & 4.0 & 103 (101) & 28.0 & 0.8 & 51 & 40 & 147 & 4.4  & 156\\
B30 & 30 & 1.3 & 28 (28) & 4.4 & 124 (124) & 46.3 & 1.2 & 76 & 43 & 141 & 3.4  & 74\\
B40 & 40 & 1.4 & 22 (22) & 4.5 & 100 (100) & 71.9 & 1.7 & 108 & 22 & 75 & 6.9 & 96\\
B50 & 50 & 1.5 & 31 (31) & 4.3 & 132 (132) & 75.8 & 1.3 & 105 & 28 & 90 & 6.3 & 83\\
\br
\endfulltable

\begin{figure}[tbh!]
 \centering
 \includegraphics[width=0.6\linewidth]{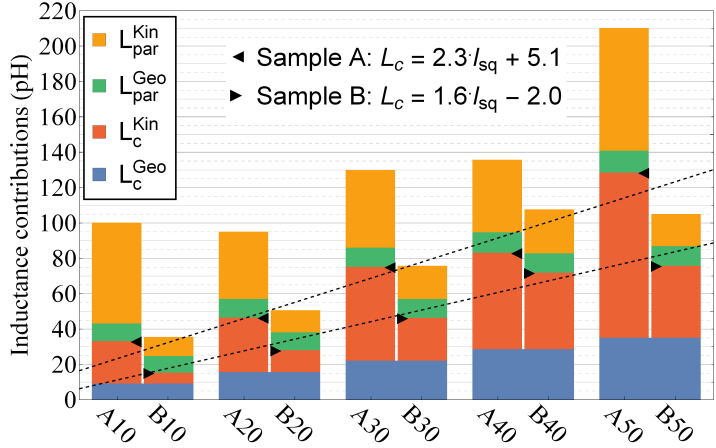}
 \caption{\label{fig:Inductance} Inductance contributions $L_c^{Geo}$, $L_c^{Kin}$, $L_{par}^{Geo}$ and $L_{par}^{Kin}$ of SQUIDs on sample A (left bar) and B (right bar) for increasing SQUID loop length l$_{sq}$. The black triangles ($\blacktriangleleft$ for sample A and $\blacktriangleright$ for sample B) mark the measured coupling inductance $L_c$. The black dashed lines are a linear fit to the measured $L_c$ for the two samples.}
\end{figure}

Results from $L_c$ measurements with direct injection for the 10 studied devices are presented as black triangles in Figure \ref{fig:Inductance}. The two black dotted lines represent separate linear fits to the data points from sample A ($\blacktriangleleft$) and sample B ($\blacktriangleright$). $L_c$ is much higher in sample A; the average inductance per unit length given by the slope of the fit is 2.3 pH/\um{} in sample A and only 1.6 pH/\um{} in sample B.
The reason for this $L_c$ discrepancy between samples is the much larger kinetic inductance contribution in sample A. To divide $L_c$ into geometric and kinetic inductance contributions, we assume that $L_c^{Geo}$ is given by the simulated value for the particular $l_{sq}$, and set $L_c^{Kin} = L_c - L_c^{Geo}$. For sample A, the kinetic inductance accounts for 66-72\% of $L_c$, while
in sample B it is 40-60\%. Hence it is not only in nanoSQUIDs where the kinetic inductance plays a significant role in the coupling between the pickup loop and the SQUID \cite{Arzeo2016,McCaughan2016,Xie2017}, but also in narrow linewidth ($\sim$4.5 \um{}) hairpin SQUID magnetometers with a directly coupled pickup loop.

By comparing the obtained value for $L_c^{Kin}$ with the simulated one, $\lambda_p$ can be extracted: $\lambda_p = \lambda_p(\text{sim.})\cdot L_c^{Kin}/L_c^{Kin}(\text{sim.})$. For sample A, large values between 2.1 \um{} and 3.0 \um{} are obtained, while $\lambda_p$ ranges between 0.7 \um{} and 1.3 \um{} for sample B. The difference between the two samples is due to the different critical temperatures, on which $\lambda$, and thus $\lambda_p$, is strongly dependent. We attribute the differences in $\lambda_p$ between SQUIDs on the same sample to differences in film thickness and $T_c$ across the sample, as well as small differences in the operation temperature during the measurements. All these factors result in strongly varying kinetic inductance contributions for different SQUIDs.

The knowledge of $\lambda_p$ allows us to estimate $L_{par}^{Kin}$ from the simulated value for the relevant geometry: $L_{par}^{Kin} = L_{par}^{Kin}(\text{sim.})\cdot \lambda_p/\lambda_p(\text{sim.})$. For $L_{par}^{Geo}$, we directly use the simulated values. The resulting $L_{par}$ is 53-82 pH in sample A and 20-36 pH in sample B (see Figure \ref{fig:Inductance}). While $L_c^{Kin}$ is not a problem as it can simply be adjusted by varying $l_{sq}$, large $L_{par}^{Kin}$ (as obtained in sample A) needs to be avoided because $L_{par}$ does not contribute to the coupling, but only to L, which reduces \Vphi{} and increases \Sphi{} (more on this in the next section).

The resulting total inductance $L$ is on average around twice as large in sample A compared to sample B. This makes it evident that coupling inductance measurements are crucial in order to estimate $L$ for such narrow linewidth hairpin SQUIDs as the kinetic inductance contribution cannot be predicted by simulations only.

\subsection{Transfer function and flux noise dependence on inductance}
\begin{figure}[bth!]
 \centering
 \includegraphics[width=0.6\linewidth]{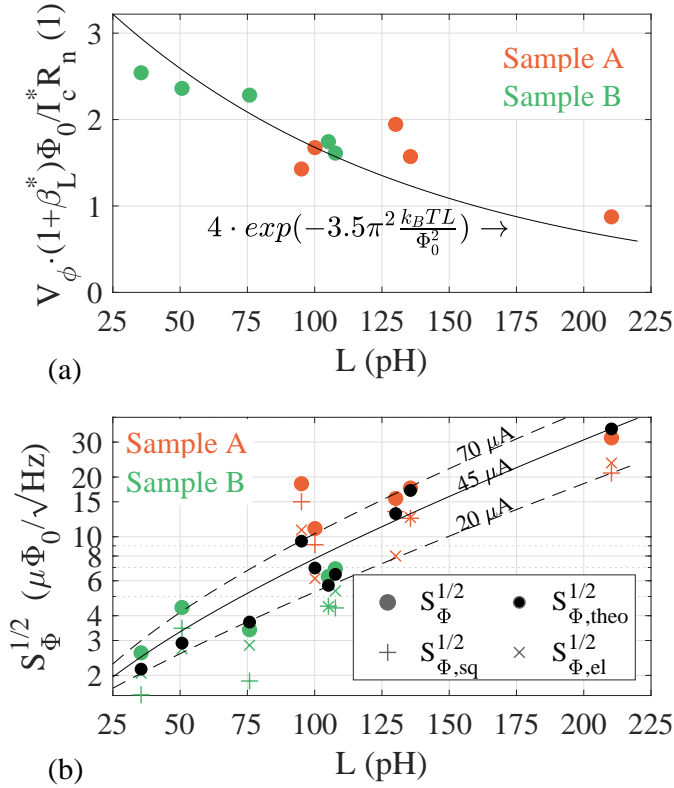}
 \caption{\label{fig:VphiSphi} (a) Normalized transfer function \Vphi{} for increasing SQUID inductance L. The black line is the prediction from equation (\ref{eq:vphi}) (i.e., without fitting parameters). (b) White flux noise levels as a function of inductance. Colored dots denote the total measured flux noise, black dots the prediction for the total flux noise based on equation (\ref{eq:sphi}), plus signs the intrinsic SQUID contribution, and crosses the flux noise due to the electronics. The dashed and solid lines are predictions based on equation (\ref{eq:sphi}) with $I_c^*R_n$ = 108 \uv{} and $I_c^*$ as indicated in the figure.}
\end{figure}

We now use the obtained $L$ to study how \Vphi{} and \Sphi{} are affected by increasing inductance.
Enpuku et al. found from simulations that the maximal \Vphi{} is determined by the expression \cite{Enpuku1994}:
\begin{equation}
V_\Phi = \frac{4}{\Phi_0}\cdot\frac{I_c^*R_n}{1+\beta_L^*}\cdot exp{\left(-3.5\pi^2\frac{k_BTL}{\Phi_0^2}\right)},
\label{eq:vphi}
\end{equation}
and hence decreases exponentially for increasing $L$ and constant $\beta_L^* = I_c^*L/\Phi_0$. As $I_c^*R_n$ and $\beta_L^*$ are not constant in our case, we show in Figure \ref{fig:VphiSphi}a the measured \Vphi{} normalized with $I_c^*R_n/\Phi_0/(1+\beta_L^*)$. The prediction of equation (\ref{eq:vphi}) with T = 78 K is shown as a solid line and describes the obtained values well -- especially in the case of sample B. Deviations may be due to resonances caused by the parasitic capacitance from the large dielectric constant of the STO substrate \cite{Enpuku1996}; we see such resonances in several of the devices at the expected voltages. Asymmetries in the SQUID parameters can also affect \Vphi{} \cite{Tesche1977}. The voltage modulation curves are furthermore not fully sinusoidal. For sinusoidal voltage modulation, the maximal \Vphi{} and the peak-to-peak voltage modulation depth $\Delta V$ are related by $\Delta V = \alpha \cdot V_{\Phi}/\pi$ with $\alpha$ = 1; for our SQUIDs $\alpha$ ranges between 0.82 and 0.98.

\begin{figure}[htb!]
 \centering
 \includegraphics[width=0.6\linewidth]{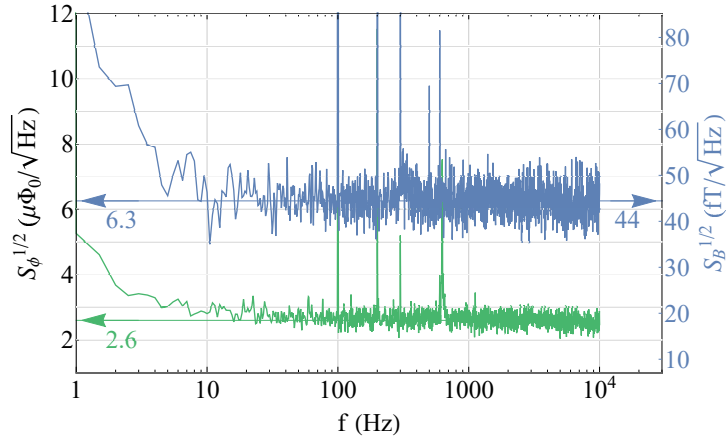}
 \caption{\label{fig:Noise} Lower curve: Total flux noise of the bare hairpin SQUID B10 as a function of frequency. This SQUID had the lowest total flux noise of 2.6 \uphihz{} (averaged between 1 and 10 kHz). Upper curve: Total flux noise and equivalent magnetic field noise of the magnetometer with the lowest equivalent magnetic field noise. The magnetometer has a 8.6 mm $\times$ 9.2 mm pickup loop (1 mm linewidth) directly coupled to the hairpin SQUID. The average noise level between 1 and 10 kHz is 6.3 \uphihz{}, corresponding to 44 fT/\rhz{}. Both measurements were done inside a superconducting shield at $T\approx$ 78 K. The devices were operated in a FLL with AC bias reversal.}
\end{figure}

The measured total (i.e., SQUID plus electronics) white flux noise levels of the 10 SQUIDs are plotted in Figure \ref{fig:VphiSphi}b as colored dots. \Sphi{} generally improves for smaller $L$ because of the decay in \Vphi{} with increasing $L$. The lowest total flux noise level of 2.6 \uphihz{} was reached for SQUID B10 with the lowest $L$ = 35 pH and the highest \Vphi{} = 195 \uvphi{}. The flux noise spectrum of this SQUID is shown in Figure \ref{fig:Noise}.

We can estimate the electronics contribution to the flux noise using $S_{\Phi,el}^{1/2} = S_{v,el}^{1/2}/V_\Phi$, where $S_{v,el}^{1/2}$ is the voltage noise from the preamplifier (0.4 nV/\rhz{} in our case). The intrinsic SQUID noise is obtained by assuming the SQUID and electronics noise are uncorrelated and that they dominate the total noise. As such, one can use $S_{\Phi,sq}^{1/2} = \sqrt{S_{\Phi}-S_{\Phi,el}}$ to extract the intrinsic SQUID noise. The two contributions are shown in Figure \ref{fig:VphiSphi}b. The dominating source of noise (be it intrinsic to the SQUID (marked with plus signs) or the electronics (marked with crosses)) depends on the individual SQUID. For the SQUID with the lowest flux noise, the electronics flux noise contribution is 2.1 \uphihz{}, resulting in an intrinsic SQUID noise contribution of only 1.5 \uphihz{} at 78 K.

To compare the obtained noise values with theory, we use that the total flux noise of a SQUID can be written as \cite{Voss1981,Blomgren2002,Faley2010}:
\begin{equation}
S_{\Phi}^{1/2} = \frac{1}{V_\Phi}\sqrt{\frac{12k_BT}{R}\left[R_{dyn}^2 + \frac{\left(LV_\Phi\right)^2}{4} \right] + S_{v,el}},
\label{eq:sphi}
\end{equation}
and assume that the normal resistance R of a single junction is given by $2R_n$, the SQUID dynamic resistance $R_{dyn}\approx \sqrt{2}R_n$, and that \Vphi{} is determined by equation (\ref{eq:vphi}). The predicted total flux noise values are indicated in Figure \ref{fig:VphiSphi}b as black dots, and fit well with the measured ones: for half of the SQUIDs the difference is less than 10\%, while on average the measured total flux noise values for all SQUIDs are 24\% larger than the predicted ones.
In Figure \ref{fig:VphiSphi}b, we plot the prediction from equation (\ref{eq:sphi}) for three different values of $I_c^*$ representing the measured range and fix $I_c^*R_n$ to the average measured value (108 \uv{}). Lower values of $I_c^*$ are clearly favourable because \Vphi{} decays with increasing $\beta_L^*$. For sample A that would mean narrower junctions with even larger parasitic kinetic inductance contributions. For all our SQUIDs the theoretical intrinsic SQUID noise is dominated by the $R_{dyn}$-term in equation (\ref{eq:sphi}), while the $L$-term is negligible.

\subsection{Magnetic field noise optimization}
Because of the trade-off between low flux noise and strong coupling, the lowest magnetic field noise is not achieved with the SQUID that has the lowest flux noise. Instead, it is the SQUID with the lowest flux noise to coupling inductance ratio $S_\Phi^{1/2}/L_c$. Of the 10 SQUIDs, B30 had the lowest $S_\Phi^{1/2}/L_c = 74$ \uphihz/nH. However, for SQUIDs with $l_{sq}$ = 30-50 \um, $S_\Phi^{1/2}/L_c$ varies only slightly with $L_c$, but also depends on $L_{par}$, $I_c^*$ and $R_n$. All these values are strongly dependent upon film quality and Josephson junction size. We therefore chose to fabricate our new magnetometers with two SQUIDs with different loop length ($l_{sq}$ = 30 \um{} and 50 \um{}) coupled to the same pickup loop. We then select the better SQUID for magnetic field detection.

Figure \ref{fig:Noise} presents the best equivalent magnetic field noise spectrum obtained with a magnetometer that had a 1 mm linewidth pickup loop as shown in Figure \ref{fig:Magnetometer}b. The SQUID with $l_{sq}$ = 50 \um{} had $L_c$ = 75 pH and $A_{eff}$ = 0.292 mm$^2$ and achieved a flux noise level of 6.3 \uphihz{}, corresponding to an equivalent magnetic field noise of 44 fT\rhz{}. With $S_\Phi^{1/2}/L_c = 84$ \uphihz/nH, the SQUID performance was slightly worse than that of B30, meaning lower magnetic field noise levels are possible if the performance of B30 can be replicated. Furthermore, the 1 mm linewidth pickup loop used here is not optimal.

The key figure describing the performance of a pickup loop is the ratio $A_p/L_p$, which should be maximized. The effective area can only be used to compare pickup loops if $L_c$ is the same, as $A_{eff}$ is strongly dependent on $L_c$. We can estimate $A_p/L_p$ from the measured $A_{eff}$ and $L_c$ using equation (\ref{eq:Aeff}), which gives $A_p/L_p$ = 3.89 mm$^2$/nH for this pickup loop. When characterizing the performance of a pickup loop this way, one must bear in mind that the value for $L_c$ is slightly overestimated because part of the injected current flows around the pickup loop instead of through the SQUID inductance. Due to the large inductance mismatch between the two, the error is less than 1\% for this pickup loop, but can become larger for pickup loops with small $L_p$.

The ratio $A_p/L_p$ is maximized for a pickup loop with a linewidth corresponding to one third of the outer pickup loop diameter $D$ \cite{Koelle1999}. We have measured $A_p/L_p$ = 5.36 mm$^2$/nH for such a pickup loop with $D$ = 9.2 mm, which based on SQUID B30 suggests that magnetic field noise levels below 30 fT/\rhz{} at 78 K are possible for the type of Josephson junction (grain boundary) and substrate size (10 mm x 10 mm) we use herein. 

\subsection{Temperature dependent magnetometer calibration}
\begin{figure}[bth!]
 \centering
 \includegraphics[width=0.6\linewidth]{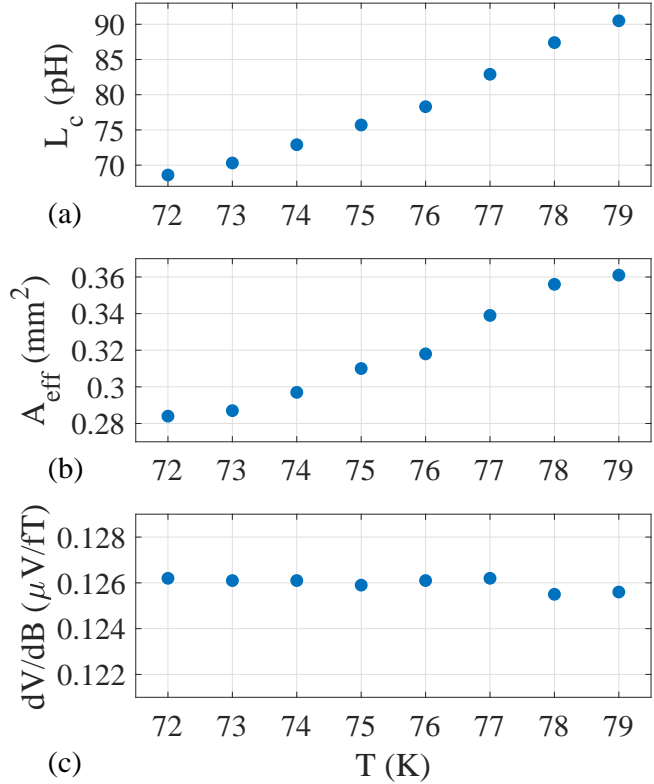}
 \caption{\label{fig:Temperature} Temperature dependence of (a) the coupling inductance, (b) the effective area, and (c) the responsivity of a hairpin SQUID magnetometer with a 1 mm linewidth directly coupled pickup loop.}
\end{figure}

Finally, we present measurements from a hairpin SQUID magnetometer operated at different temperatures (72-79 K) inside a liquid nitrogen cryostat. Decreasing the operation temperature by pumping on the liquid nitrogen bath is a simple and effective way to increase $I_cR_n$ and $\Delta V$, which normally leads to better device performance \cite{Hato2019}. For the magnetometer presented here, $\Delta V$ increases from 13 \uv{} at 79 K to 33 \uv{} at 72 K.
Figure \ref{fig:Temperature} shows that $L_c$ and $A_{eff}$ also vary strongly with temperature, even in this limited temperature range. The coupling inductance of the SQUID ($l_{sq}$ = 50 \um) drops from 91 pH to 69 pH because $L_c^{Kin}$ decreases. The effective area of the 1 mm linewidth pickup loop magnetometer follows accordingly: it decreases from 0.361 mm$^2$ to 0.284 mm$^2$. This 21\% decrease clearly demonstrates that $A_{eff}$ can not be measured at one temperature and used for sensor calibration at other temperatures.

Nonetheless, we found that by using direct injection of current as the feedback method \cite{Ruffieux2017}, the responsivity $dV/dB$ of a magnetometer operated in a flux-locked loop (FLL) with direct readout can be made temperature independent. The transfer function of a SQUID operated in a FLL with direct readout is $V_\Phi^{FLL} = R_f/M_f$, where $R_f$ is the feedback resistance (30 k$\Omega$ for our electronics) and $M_f$ the mutual inductance between the feedback coil and the SQUID loop \cite{Drung2003}. In the case of direct injection feedback (wiring as shown in Figure \ref{fig:Magnetometer}a), $M_f$ is the same as $L_c$ (note that again a small part of the injected current goes through the pickup loop). The responsivity $dV/dB$ is then given by $dV/dB = V_\Phi^{FLL}\cdot A_{eff} \approx R_f \cdot A_p/L_p$ (see equation (\ref{eq:Aeff})), and is hence independent of $L_c$ and temperature. With the same procedure as that which we used for $A_{eff}$, $dV/dB$ was measured by varying the amplitude of a field applied with a calibrated Helmholtz coil and linearly fitting the voltage response of the SQUID. This measurement gives $dV/dB$ directly and $V_\Phi^{FLL}$ does not need to be known (as it does when measuring $A_{eff}$). Figure \ref{fig:Temperature} shows that $dV/dB$ varies by less than 0.4\% in the measured temperature range, thus making reliable one-time magnetometer calibration possible.

\section{Discussion}
This SQUID inductance study demonstrates that calculating $L$ numerically is not enough to determine $L$ for narrow linewidth SQUIDs as the kinetic inductance contribution can vary greatly between different samples, hence making inductance measurements necessary. Direct injection of current into the SQUID loop is a very simple method to measure inductance and can thus easily be integrated into the standard SQUID characterization routine. Herein, we demonstrated a number of additional benefits of direct injection of current into the SQUID loop.  First, it can be used to couple flux into the SQUID loop, thus eliminating the need for additional coils for SQUID tuning and feedback \cite{Ruffieux2017}. Second, the demonstrated temperature independence of the sensor's responsivity ($dV/dB$) with this approach allows confidence in the calibration even when the temperature of the sensor varies by several kelvin. That the kinetic inductance contribution decreases with decreasing temperature can be an advantage as it compensates partly for the increasing $I_c^*$ (which increases $\beta_L^*$ for lower temperatures) meaning that the SQUID can be operated in a wider temperature range. Third, the inductance can be measured at lower temperatures (i.e., also outside the temperature range of interest for a given application). This allows extraction of $\lambda_0$, verification of equation (\ref{eq:lambda}), and estimation of the individual sensor's $T_c$ \cite{Terai1997,Li2019}, which we show can vary within a single sample. 

We find good agreement between theoretical estimations and measured values both for \Vphi{} and \Sphi{} in contrast to earlier reports \cite{Enpuku1995,Koelle1999,Blomgren2002,Lam2013}. This suggests that equations (\ref{eq:vphi}) and (\ref{eq:sphi}) can be used to optimize SQUID parameters. Equation (\ref{eq:sphi}) predicts that lower noise levels can be achieved with junction technologies offering higher $I_c^*R_n$ at 77 K, such as step edge junctions \cite{Mitchell2010,Faley2014,Kaczmarek2018} or possibly the novel grooved Dayem nanobridges \cite{Trabaldo2019}.
For fixed values of $I_c^*$ and $R_n$, SQUIDs with low $L_{par}$ have lower \Sb{}, while higher values of $L_{par}$ also demand higher values of $L_c$ to minimize \Sb{}($L_c$). In order to achieve low $L_{par}$ for a fixed junction cross section, short bridges forming the junctions and high quality films are necessary. Nonetheless, it can be favourable to reduce the junction cross section in order to achieve higher $R_n$ and lower $I_c^*$ values, despite the higher $L_{par}$ associated with narrow bridges.

The findings in this paper can be summarized into a recipe for producing low magnetic field noise YBCO SQUID magnetometers with a directly coupled pickup loop. However, many dependencies of $I_c^*$, $R_n$, and $L$ have to be taken into account, and the optimal values depend on the operation temperature and the junction technology used. To limit the parameter space, we assume that the Josephson junction technology is predetermined and the operation temperature is set by the application (and all characterization measurements are done at this temperature). We furthermore assume that the film thickness has been decided upon based on the junction technology used and fabrication limits.
As a first step, the YBCO film quality needs to be optimized for low $\lambda_p$ given by a high $T_c$, and the minimal bridge length required for good Josephson junctions needs to be identified -- both in order to minimize $L_{par}^{Kin}$.
Next, the optimal SQUID design parameters $w_{JJ}$ and $l_{sq}$ defining $I_c^*$, $R_n$, and $L$ need to be determined. To this end, the dependencies of $I_c^*$ and $R_n$ on $w_{JJ}$ have to be established from measurements of test SQUIDs, and $\lambda_p$ of the film needs to be measured. The two dependencies $I_c^*(w_{JJ})$ and $R_n(w_{JJ})$ can then be combined with simulated data for $L(w_{JJ},l_{sq})$ using the measured $\lambda_p$ (like in Figure \ref{fig:Simulation}) to perform a 2 dimensional minimization of \Sphi{}$/L_c$ based on equation (\ref{eq:sphi}), where $w_{JJ}$ and $l_{sq}$ are varied. The minimum defines the optimal values $w_{JJ}^{opt}$ and $l_{sq}^{opt}$.
The last step is to fabricate a complete magnetometer where a SQUID with the optimal SQUID dimensions $w_{JJ}^{opt}$ and $l_{sq}^{opt}$ is coupled to a wide linewidth pickup loop (1/3 of the outer pickup loop dimension \cite{Koelle1999}).

This type of optimization requires good knowledge of $I_c^*$ and $R_n$ as a function of junction dimensions and the ability to reproducibly fabricate junctions with the selected parameters, which is problematic for the grain boundary Josephson junction technology we use. This is why we fabricate each magnetometer chip with two redundant SQUIDs that not only have different inductances, but also different $w_{JJ}$ as this approach increases the likelihood that one of the two SQUIDs will be close to optimal. It is possible to fabricate even more redundant SQUIDs, but such an approach results in a reduction in the size of the pickup loop. To increase the chance that the coupled SQUID has the designed $I_c^*$, $R_n$, and $L$, the same film can be used to first fabricate the test SQUIDs (e.g., in the center of the chip) and later fabricate the complete magnetometer. Another option is to trim the junction with ion beam milling if the initial $I_c^*$ is too high in both SQUIDs \cite{Du2001}.

The variable kinetic inductance contributions discussed in this paper are not only important for bare hairpin SQUIDs and hairpin SQUID magnetometers with a directly coupled pickup loop, but also for other devices that contain narrow YBCO lines and are operated close to $T_c$. Examples of such devices include hairpin SQUID gradiometers with a directly coupled pickup loop, washer SQUIDs with holes or slots \cite{Dantsker1997}, SQUID arrays \cite{Chesca2015,Taylor2016}, and superconducting quantum interference filters (SQIFs) \cite{Taylor2016,Mitchell2016,Coueedo2019}. Depending on the superconductor used and how close $T_c$ is to the operation temperature, variations in the kinetic inductance contributions can also be expected in nanoSQUIDs \cite{Granata2016} and superconducting digital logic circuits \cite{Przybysz2015}.

\section{Conclusion}
We performed an inductance study to optimize the noise levels of our YBCO hairpin SQUIDs and their coupling to a directly coupled pickup loop.
By combining inductance simulations and coupling inductance measurements, we could differentiate between the kinetic and the geometric inductance contributions as well as extract $L_c$, $L_{par}$ and $L$.
We found that the kinetic inductance plays an important role as it comprises a significant contribution to the total inductance in these 4.5 \um{} linewidth SQUIDs and varies both with film quality and temperature.
A comparison between two samples with bulk critical temperatures of 87.7 K and 89.0 K revealed that $L$ can differ by a factor of 2 for the same SQUID loop size, hence making inductance measurements a crucial part of SQUID characterization and optimization.
We furthermore found good agreement between measured values and theoretical estimates for \Vphi{} and \Sphi{}, which allows optimization of SQUID sensor performance. The lowest total flux noise level reached with a bare SQUID at 78 K was 2.6 \uphihz{}. The magnetometer with the lowest magnetic field noise level at 78 K achieved 44 fT/\rhz{}.
Finally, we demonstrated a method for reliable magnetometer calibration despite temperature dependent coupling.
The presented inductance study provides a wealth of insights into the design, characterization, optimization, and operation of narrow linewidth YBCO SQUID-based devices operated close to $T_c$.

\section*{Acknowledgments}
This work is financially supported by the Knut and Alice Wallenberg foundation (KAW 2014.0102), the Swedish Research Council (621-2012-3673), the Swedish Childhood Cancer Foundation (MT2014-0007), Tillv\"axtverket via the European Regional Development Fund (20201637), and the ATTRACT project funded by the European Commission (777222). We also acknowledge support for device fabrication from the Swedish national research infrastructure for micro and nano fabrication (Myfab).

\section*{References}
\bibliographystyle{iopart-num}
\bibliography{InductanceReferencesArxiv}
\end{document}